# Hollow chain-like beams

Dmitriy.Yu. Cherepko[a], Nataliya D. Kundikova[a,b], Ivan I. Popkov[a,b]

[a]*Department of Optics and Spectroscopy, South Ural State University*
*76 Lenin Av., Chelyabinsk 454080, Russia*
[b] *Nonlinear Optics Laboratory, Institute of Electrophysics, RAS,*
*76 Lenin Av., Chelyabinsk 454080, Russia*

**Abstract**

To generate hollow chain-like beams the diffraction of the first order Bessel beam by zone plate with two odd open Fresnel zone has been investigated. It has been shown that the capsules size is influenced by the number of the second odd open Fresnel zone and by the zone plate focal length. A hollow chain-like beam has been experimentally generated as a result of the first order Bessel beam diffraction by zone plate with the first and the ninth open Fresnel zones. The orbital angular momentum presence has been proved experimentally. The main features of the beam have been investigated. Sufficiently good agreement between experimental and numerically calculated results has been demonstrated.

## 1. Introduction

Last decades, beams with different intensity and phase space structure are used for an optical microparticles manipulation instead of a focused Gaussian beam. Laguerre-Gaussian beams and Bessel beams of the first order are the example of such beams. Due to an orbitalangular momentum it is possible not only to capture but also rotate microparticles [1, 2, 3, 4]. Such beams have opened the way to a simple technique for atom manipulation [3, 4]. Simultaneous tweezing of low index particles and high index particles in the axial direction is possible also [5].

The beams having self-similar structure of intensity distribution along a direction of propagation are the other class of beams which are used for opti-cal microparticles manipulations. Such beams allow to capture and operate group of particles and tweezing of low index particles [6, 7, 8]. The chain-like beam is a result of Gaussian beam diffraction by a zone plate with several open odd zones [6, 7]. A diffraction pattern of a Gaussian beam passing through a fractal zone plate is a light beam with self-like intensity structure along the beam propagation direction [8]. The usage of generalized zone plates allows to increase the number of focal points, saving property of self-similarity [9]. The main difference between that beams [8, 9] and a chain-like beams [6, 7] is intensity distribution between principal and secondary focal points and the number of focal points [6, 7, 8, 9]. Another example of the beams with nonuniform intensity distribution along the direction of the propagation is a beam obtained by means of so called "devil's lenses" in which the phase distribution is characterized by the "devil's staircase" function [10, 11].

The combination of properties of a Bessel beam of the first order and a chain-like beam will result in a new type of hollow beams. It can be useful for various applications, in particular, for atoms manipulation and for determination of microviscosity distribution.

Beams with nonuniform intensity distribution along the propagation direction and a sequence of focused optical vortices along the propagation direction can be generated by using a spiral fractal zone plate [12]. Such zone plates can be both circular [12], and square [13, 14]. For a circular zone plate the transverse intensity distribution is a number of dark and light rings [12] as for a square zone plate the central dark spot is in the form of a rhomb [13, 14]. The combination of a Devil's lens [10, 11] and a helical vortex phase mask [15] has been also used [16].

It is possible to obtain a hollow beam with nonuniform intensity distribution along the propagation direction by means of illumination of a zone plate with a beam carrying topological charge instead of the spiral zone plate usage.

In this paper, we are going to investigate the properties of a hollow chain-like beam generated by means of the first order Bessel beam diffraction by an amplitude binary zone plate with two open odd Fresnel zones.

## 2. Numerical simulation of a hollow chain-like beam properties

We have investigated diffraction of the first order Bessel beam by an amplitude zone plate with two open odd Fresnel zones. The paraxial approximation has been used for numerical simulation. The parabolic equation has been solved by the spectral method, based on the two-dimensional Fourier transform. The wavelength of radiation diffracting on the zone plate was $\lambda = 632.8$ nm, two odd Fresnel zones were open on the zone plate. The radius $R_x$ of Fresnel zone with the number $m$ is:

$$R_m = \sqrt{mF\lambda}, \qquad (1)$$

here $F$ is the zone plate focal length. The diffraction was calculated for a binary amplitude mask with the first Fresnel zone radius $R_1 = 0.96$ mm, the main focus at the distance $z_1 = R_1^2/\lambda = F = 145$ cm, and additional focal spots at the distances $z_2 = F/3 = 48$ cm and $z_3 = F/5 = 29$. Fig. 1 shows the mask with the first and ninth open Fresnel zones. The same binary amplitude mask was used in [6, 7].

Diffraction of a Gauss beam by the zone plate, depicted in Fig. 1, results in a chain of the light capsules forming chain-like beam. Diffraction of the first order Bessel by the zone plate results in a hollow chain-like beam. Diffraction trees for the first and the second cases are presented in Fig. 2.

One can see from Fig. 2 that the main difference between two beams is the presence of a dark channel at the beam axis along the propagation direction.

Numerical simulation of the first order Bessel beam diffraction by zone plates with different odd open Fresnel zones has been carried out. Fig. 3 shows the first order Bessel beam diffraction trees for the zone plate with the first and the fifth open Fresnel zones, for the zone plate with the first and the ninth open Fresnel zones and for the zone plate with the first and the thirteenth open Fresnel zones. The Bessel beam parameters were changed in order to allow the equal energy amount to be transmitted through the each zone.

It can be seen from Fig. 3 that the number of chosen zones influences the capsule longitudinal and transversal size; namely, the larger the difference between the zones numbers the smaller the capsule size. Taking into consideration our experimental set up we have carried out all following numerical simulation with the zone plate with the first and the ninth open Fresnel zones.

The relative amount of energy transmitted through the open zones of a binary amplitude mask depends on the beam width. It has been demonstrated numerically and experimentally that a decrease in the Gaussian beam width increases the capsule sizes and the focus depth, decreases the intensity in the focuses, and does not affect the capsule position [7]. The relative portions of energy which pass through two open Fresnel zones of the zone plate are different for the Gaussian beam and the first order Bessel beam. Let us consider how the first order Bessel beam width influences the beam transverse and longitudinal intensity distribution. It can be seen from Fig. 4 that change of the first order Bessel beam width results in the change of the relative portions of energy which pass through two open Fresnel zones. If the radius of the first ring of the first order Bessel beam is equal to 2.3 mm (Fig. 4a), the value of the energy transmitted through the ninth zone relative to the energy transmitted through the first zone $J_9/J_1 = 0$ will be equal to zero, if the radius of the first ring of the first order Bessel beam is equal to 3.45 mm (Fig. 4a) and 4.6 mm (Fig. 4a) the value of the relative energy will be $J_9/J_1 = 0.5$ and $J_9/J_1 = 2.5$.

Figure 5 shows the first order Bessel beam diffraction trees for the beam with the radius of the first ring equal to 2.3 mm (Fig.5a), 3.45 mm (Fig. 5b) and 4.6 mm (Fig. 5c). One can see from Fig. 5 that the change of the first ring radius of a first order Bessel beam, diffracted by the zone plate, results in the change of the contrast between the dark and the light areas of the hollow chain-like beam. If the Bessel beam illuminates only the first Fresnel zone the diffraction tree will be similar to the diffraction tree of the first order Bessel beam diffracted by a circular aperture. Light capsules are appeared under increasing of the first ring radius of the Bessel beam. If the first and the ninth Fresnel zones are illuminated with equal intensity, the contrast between dark regions and the light capsules will be maximal.

The beam profiles at the distance 1.26 m for two beams whose diffraction trees are depicted in Fig. 5b and Fig. 5c are presented in Fig. 6. One can see from Fig. 6 that increase of the value $J_9/J_1$ results in the contrast decreasing.

The influence of the zone plate focal length on the beam diffraction tree has been investigated. The radii of the first and the ninth zone have been changed according to Eq. (1), the Bessel beam parameters have been changed in such a way to keep the value $J_9/J_1$ unchanged and equal to 1. The results of computer simulation are presented in Fig. 7. The dashed line marks the distance of 1 meter from the zone plate.

One can see from Fig. 7 that the increase of the zone plate focal lens results in the change of the capsules size and their positions, but the main beam features remain unchanged. It means that we can control capsules size and their position simply by changing the zone plate focal length. The computer simulation has allowed us to choose the parameter of the zone plate and the Bessel beam to carry out experimental investigation.

## 3. Experimental generation of the hollow chain-like beam

The experimental setup designed to investigate the properties of the hollow chain-like beams is shown in Fig. 8.

The key elements of the setup are two amplitude masks. The first one is an amplitude mask (the Bessel mask) obtained by the following way. The interference pattern of a Gaussian beam and the first order Bessel beam has been generated on the computer and negative image of the interference pattern has been printed in enlarged size. The printed negative image has been photographed using a film with high resolution. That film $3.5 \times 2.5$ cm$^2$ in size has been used to generate the first order Bessel beam. The second one is a binary amplitude mask $3.5 \times 2.5$ cm$^2$ in size with the first and ninth open Fresnel zones and the first Fresnel zone radius $R_1 = 0.96$ mm, generated on the computer ($256 \times 256$ pixels) and printed on a transparency with resolution 600 dpi (Fig. 2). The output of a He-Ne laser with the wavelength $\lambda = 632.8$ nm and power of 1.5 mW has been used in the experiment.

The laser beam has been expanded by an optical system to a width of 3 cm. The expanded collimated coherent monochromatic light has been interacted with the Bessel mask and the diffracted pattern has been focused in order to choose only the desired order from the diffraction pattern in the focal plane by diaphragm. The selected first order Bessel beam has been collimated and interacted with the second amplitude mask. The value $J_9/J_1$ was equal to 7 in the experiment. The intensity distribution across the transverse section of the resulting hollow chain-like beam has been registered by a CCD camera at the different propagation distances. The CCD matrix (VEC-545) had a registration zone of $2592 \times 1944$ pixels ($5.808 \times 4.294$ mm$^2$), each pixel of about 2.2 $\mu$m.

The beam transverse intensity distribution at the distance 1.6 m from the mask is shown in Fig. 9. The left part of the image is the experimental image and the right part is the calculated image at the same distance from the mask and $J_9/J_1 = 7$.

From Fig. 9 one can see that there is reasonably good agreement between the experimental and the calculated transverse intensity distribution.

In order to check the orbital angular momentum presence in the beam under investigation additional optical elements forming interferometer have been used. The laser beam has been divided into two beams, the first have been used to obtain the beam under investigation and the second one, expanded and collimated, has interfered with the beam under investigation. The interference pattern of the Gaussian beam and the hollow chain-like beam is shown in Fig. 10. The interference patterns in Fig. 10a and Fig. 10b are the same, but a dash line at the Fig. 10b shows a "fork" position.

A fork-like dislocation (Fig. 10) with a difference of one arm corresponds to phase winding by $2\pi$ around the beam core and proves the orbital angular momentum presence in the beam.

Using a CCD matrix the transverse intensity distribution of the hollow chain-like beam has been recorded at distances ranged from 100 to 201 cm with a step of 1 cm. The experimental intensity distributions have been processed to perform the diffraction tree in the following way. A section fitting the length of the beam diameter has been selected from every intensity distribution, so that the thickness of each section corresponds to one pixel of the CCD camera (2.2 $\mu$m in the actual case). Then, placing the sections consecutively in growing order, the image representing a part of the diffraction tree of the beam under investigation has been build up. The left part of Fig. 11 presents the processed experimental data. The right part of Fig. 11 shows the diffraction tree calculated numerically under the same parameters. It can be easily seen from Fig. 11 that the diffraction tree, constructed from the experimental data, depicts the main features of the diffraction tree calculated numerically. The whole beam structure resembles the chain-like beam with the dark channel at the beam axis.

In summary, the diffraction of the first order Bessel beam by zone plates with two odd open Fresnel zone has been investigated. The result of the diffraction is hollow chain-like beam. It has been shown that the number of the second odd open Fresnel zone influences the capsule longitudinal and transversal size; namely, the larger the difference between the zones numbers the smaller the capsule size is.

The hollow chain-like beam has been experimentally generated as a result of the first order Bessel beam diffraction by zone plate with the first and the ninth open Fresnel zones. The orbital angular momentum presence has been proved experimentally. The main features of the beam have been investigated. Sufficiently good agreement between experimental and numerically calculated results has been demonstrated.

**List of Figure Captions**

Fig. 1. Fig. 1.The image of the binary amplitude mask with the first and ninth open Fresnel zones, $F = 145$ cm.

Fig. 2. Diffraction trees for the zone plate with the first and ninth open Fresnel zones, a) Gaussian beam diffracts, b) the first order Bessel beam diffracts.

Fig. 3. The first order Bessel beam diffraction trees for the zone plate with the first and the fifth open Fresnel zones (a), for the zone plate with the first and the ninth open Fresnel zones (b) and for the zone plate with the first and the thirteenth open Fresnel zones (c). The equal amount of the light energy was transmitted through two open zones in each case.

Fig. 4. The illumination of the zone plates with the first and the ninth open Fresnel zones with the first order Bessel beam, a)$J_9/J_1 = 0$, b)$J_9/J_1 = 0.5$ and c)$J_9/J_1 = 2.5$.

Fig. 5. The first order Bessel beam diffraction trees for the beam with the radius of the first ring equal to 2.3 mm (a), 3.45 mm (b) and 4.6 mm (c).

Fig. 6. The profile of the chain-like beam at the distance 1.26 m for the different radius of the first Bessel beam ring: a)$J_9/J_1 = 0.5$, b)$J_9/J_1 = 2.5$.

Fig. 7. The influence of the zone plate focal length on the beam diffraction tree, $J_9/J_1 = 1$, a)$F = 0.9$, b)$F = 1.2$, c)$F = 1.5$.

Fig. 8. The experimental set-up used to investigate the properties of the. More details are in the text.

Fig. 9. The experimental (left) and calculated (right) beam transverse intensity distribution at the distance 1.6 m, $J_9/J_1 = 7$.

Fig. 10. The interference pattern of the Gaussian beam and the chain-like beam with the phase singularity.

Fig. 11. The diffraction tree, constructed from the experimental data (left), and calculated numerically (right); $F = 1.4$ m, $J_9/J_1 = 7$.

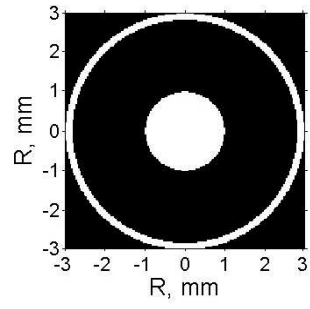

Figure 1: The image of the binary amplitude mask with the first and ninth open Fresnel zones, $F = 145$ cm.

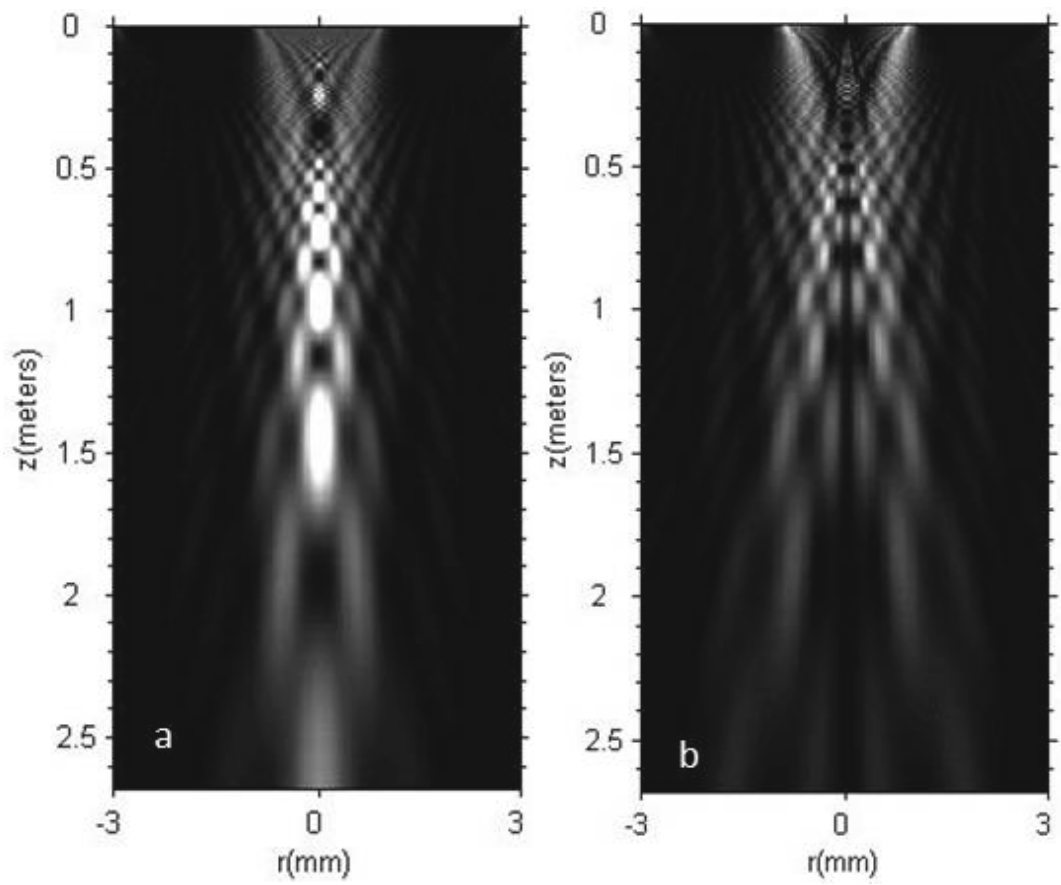

Figure 2: Diffraction trees for the zone plate with the first and ninth open Fresnel zones, a) Gaussian beam diffracts, b) the first order Bessel beam diffracts.

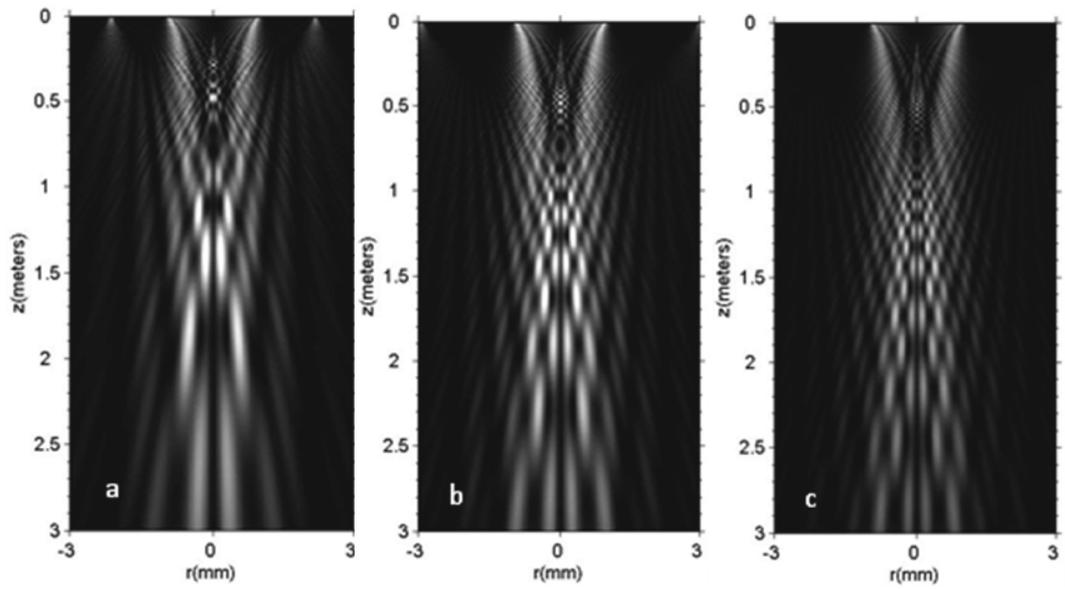

Figure 3: The first order Bessel beam diffraction trees for the zone plate with the first and the fifth open Fresnel zones (a), for the zone plate with the first and the ninth open Fresnel zones (b) and for the zone plate with the first and the thirteenth open Fresnel zones (c). The equal amount of the light energy was transmitted through two open zones in each case.

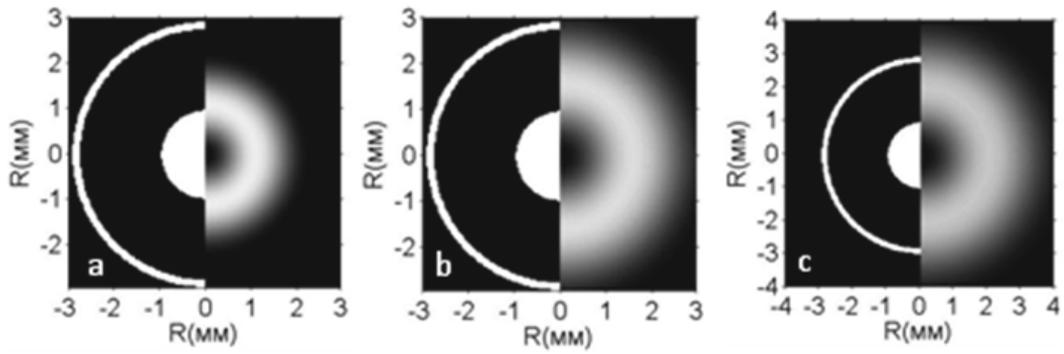

Figure 4: The illumination of the zone plates with the first and the ninth open Fresnel zones with the first order Bessel beam, a)$J_9/J_1 = 0$, b)$J_9/J_1 = 0.5$ and c)$J_9/J_1 = 2.5$.

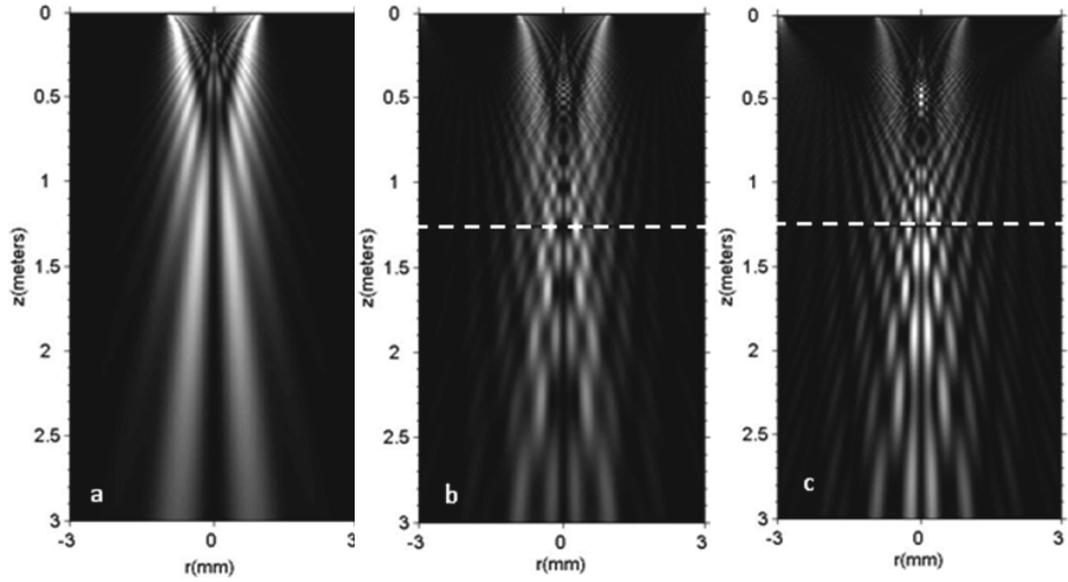

Figure 5: The first order Bessel beam diffraction trees for the beam with the radius of the first ring equal to 2.3 mm (a), 3.45 mm (b) and 4.6 mm (c).

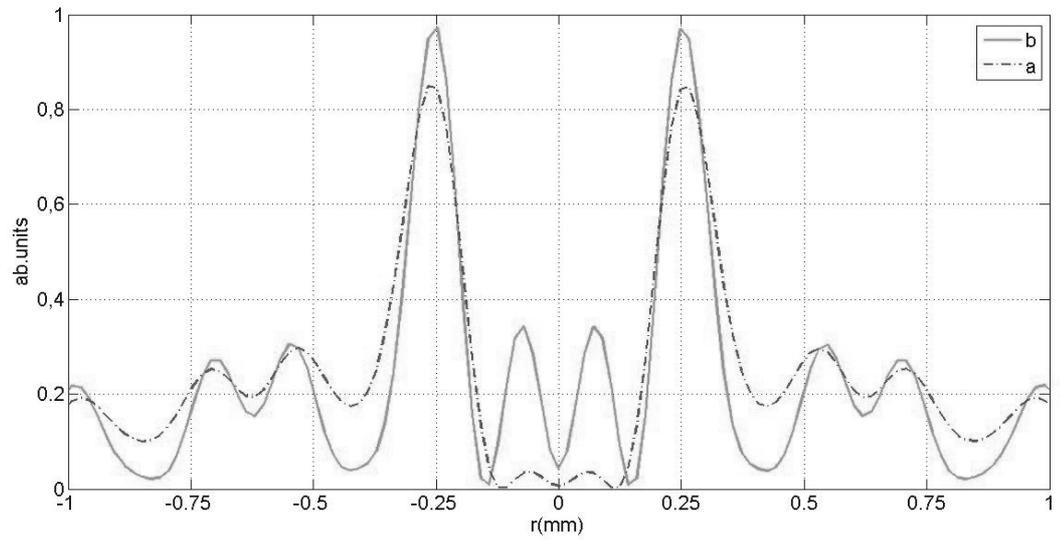

Figure 6: The profile of the chain-like beam at the distance 1.26 m for the different radius of the first Bessel beam ring: a)$J_9/J_1 = 0.5$, b)$J_9/J_1 = 2.5$.

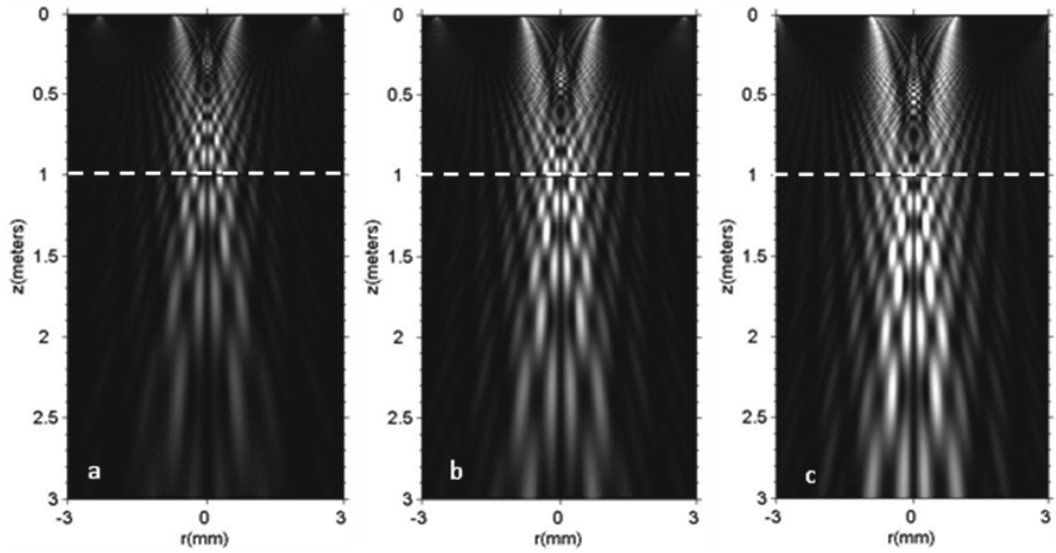

Figure 7: The influence of the zone plate focal length on the beam diffraction tree, $J_9/J_1 = 1$, a)$F = 0.9$, b)$F = 1.2$, c)$F = 1.5$.

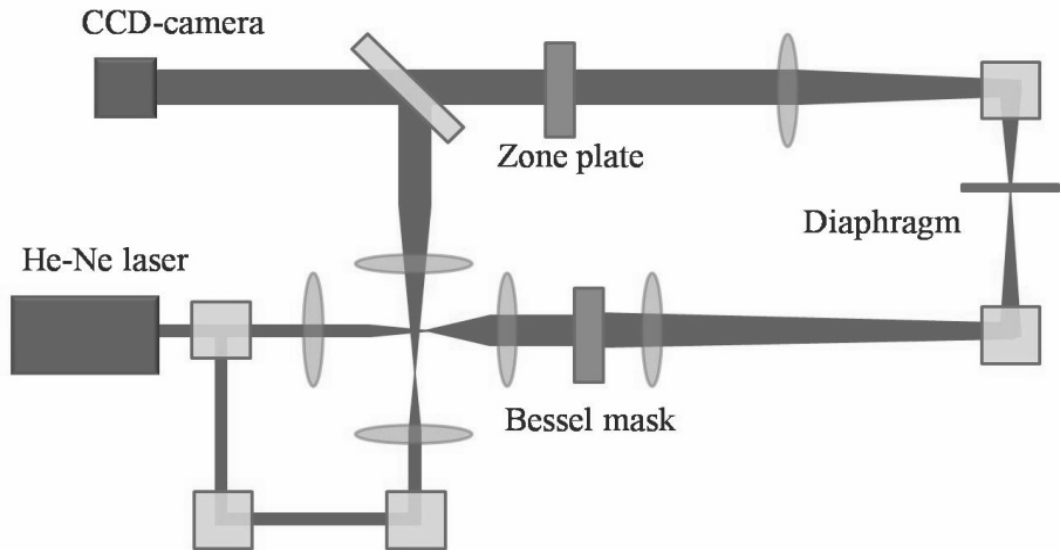

Figure 8: The experimental set-up used to investigate the properties of the. More details are in the text.

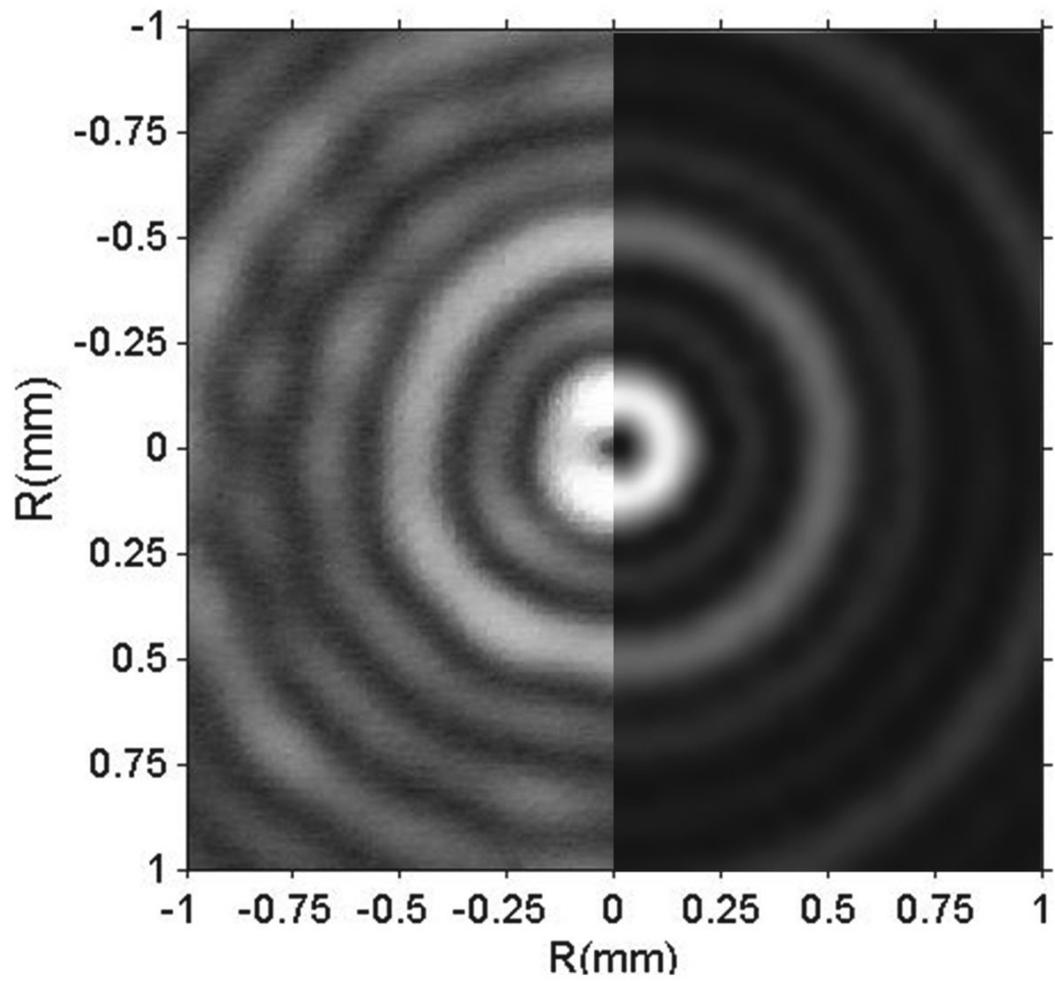

Figure 9: The experimental (left) and calculated (right) beam transverse intensity distribution at the distance 1.6 m, $J_9/J_1 = 7$.

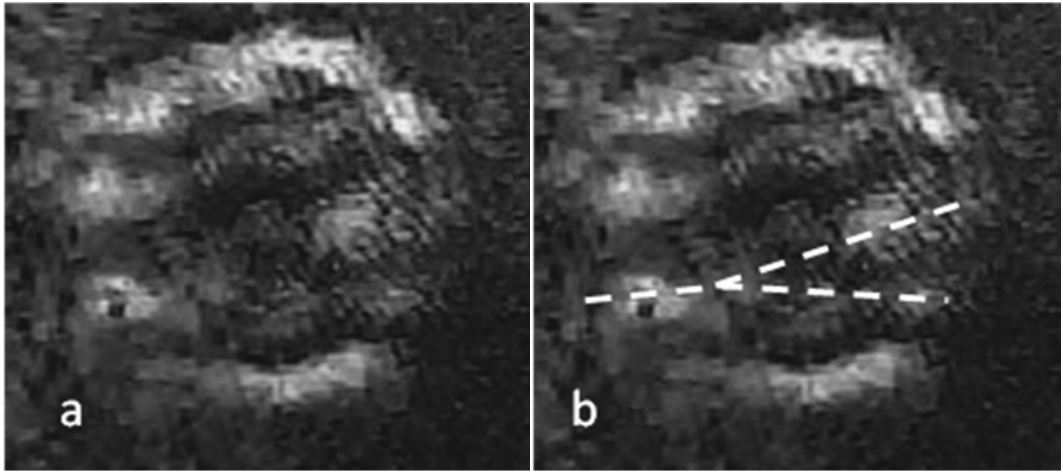

Figure 10: The interference pattern of the Gaussian beam and the chain-like beam with the phase singularity.

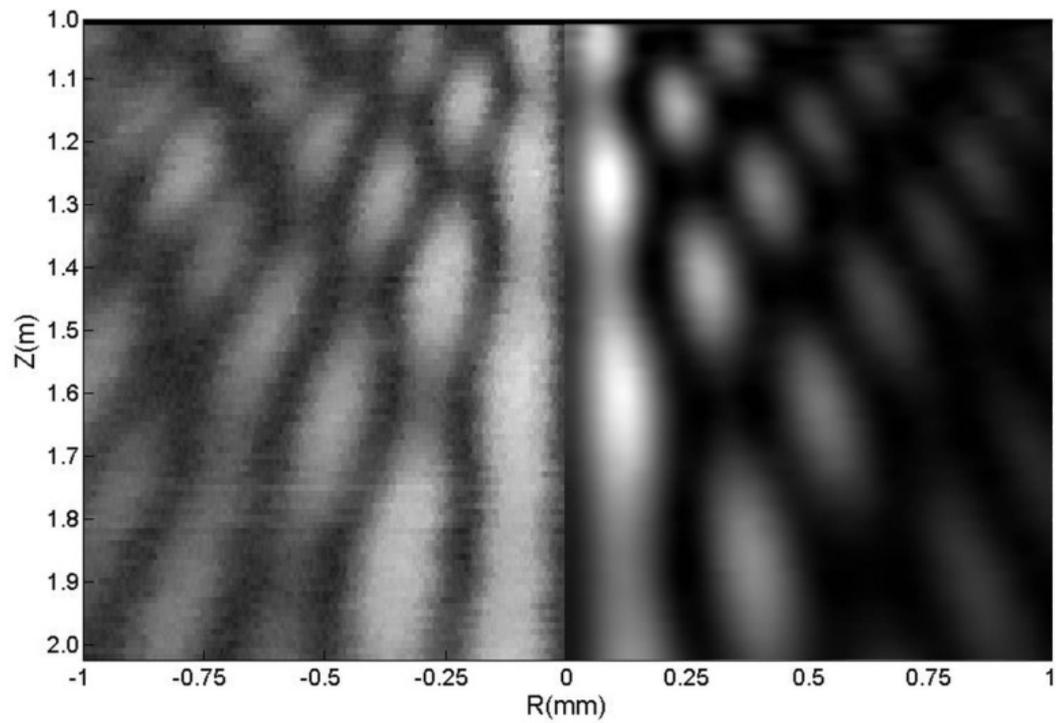

Figure 11: The diffraction tree, constructed from the experimental data (left), and calculated numerically (right); $F = 1.4$ m, $J_9/J_1 = 7$.